\documentclass[11pt,twoside]{article}
\usepackage{asp2010,graphicx,setspace,natbib}

\resetcounters

\begin{document}

\title{Creation of Peanut-Shaped Bulges via the Slow Mode of Bar Growth}
\author{Michael S. Petersen,$^1$ Martin D. Weinberg,$^1$ and Neal Katz$^1$
\affil{$^1$University of Massachusetts at Amherst}
}

\begin{abstract}
Recent theoretical work has implicated fast bar formation modes and subsequent
evolution as the creation
mechanism for the observed peanut-shaped bulges in
some edge-on disk
galaxies. We demonstrate an 
N-body simulation of a disk undergoing a contrasting slow mode of
bar growth, unsubjected to a buckling instability, which
nonetheless grows the 4:1 orbit family responsible for a peanut-shaped
bulge. We also present a simulation with fast mode
bar growth, which exhibits thickening similar to other work. A novel
orbit classification method that finds dynamically distinct families
is presented, allowing for a detailed analysis
of angular momentum transfer channels within the disk.
\end{abstract}

\section{Introduction}

When viewed edge-on, 45\% of disk galaxies show a boxy or peanut-shaped
bulge \citep{bureau05}. Kinematic observations have revealed cylindrical rotation in some
 \citep{williams11}, which is interpreted as evidence for secular
evolution: the cylindrical rotation in these bulges is a kinematic memory
of the disk that the stars were elevated out of over many dynamical times. Boxy and
peanut bulges, therefore, stand apart from
spheroidal bulges, thought to be formed by merging processes, and hence do
not result in cylindrical rotatation.

Informative simulations of disk and halo systems have suggested boxy and peanut-shaped bulges to be
edge-on bars. Bars in these
simulations are overwhelmingly
produced by simulating an initially highly unstable disk--the fast mode of bar
formation (e.g. \citet{athanassoula02}, \citet{debattista06}, \citet{saha10}). In this mode of formation,
bars typically buckle via the firehose instability to form thickened structures. The slow mode of bar
formation \citep{polyachenko96} has been underreported, perhaps simply
owing to the uncertain nature of N-body initial conditions that would
lead to this mode. In the slow mode, angular momentum transfer
exploits the dark halo as an accepting angular momentum
reservoir for stability against
buckling instabilities. 

Identifying and isolating the channels
for angular momentum transfer
is therefore of the utmost importance for understanding the mechanisms
behind the
observed morphology. The analysis of
properly executed simulations can shed light on
the situation. In the next section, we detail two such simulations. In the third
section we discuss the analysis and dynamical implications. We conclude with a statement of the utility of this study
for informing future work.

\section{Simulations}
Previous work has focused on two general classes of initial conditions: halo- or disk-dominated
systems with cored halos. For our simulations, we select a different approach. The first simulation features a
cuspy central profile, $\rho_h \propto r^{-1}$, while the second
simulation features a cored central profile, $\rho_h\propto {\rm
  constant}$ ($R_{\rm core} = 0.01$). The rotation curves are shown in
Figure 1. In appearance, these simulations probe the same halo-
and disk-dominated parameter space; in practice, these simulations probe
different dynamical modes. The cuspy profile features a rich
spectrum of orbital frequencies as $r\to 0$, while the cored profile
allows orbits to pile up at select frequencies. This difference proves
crucial for the bar formation mechanism and subsequent evolution.

\citet{weinberg99} presented an algorithm to solve the Poisson equation
through an expansion in empirical orthogonal functions. Implemented as
EXP, a massively parallelized N-body code, such a basis
may be tailored to the length scales and asymmetries of interest,
reducing the overall degrees of freedom and subsequently the diffusive
relaxation. Our simulations follow a disk ($N_{\rm
  disk}=10^6,~a=0.01,~h=0.001,~M_{\rm disk}=0.02
M_{\rm halo}$) embedded in a live
dark matter halo ($N_{\rm halo}=10^7,~c=15$) from $T=0.0\to2.0$ (all
units are listed in virial units for scaling purposes; for a
Milky Way size galaxy, $r=1.0\to 300 {\rm kpc}$,
$M=1.3\times10^{12}M_{\rm \odot}$, $v=1.0\to 135 {\rm~ km~s^{-1}}$, $T=1.0\to 2.2~
{\rm Gyr}$). For a more detailed
discussion of initial conditions, including the distribution function realization, see \citet{holleyb05}.
\begin{figure}[h]

\centering
\includegraphics[width=0.6\textwidth]{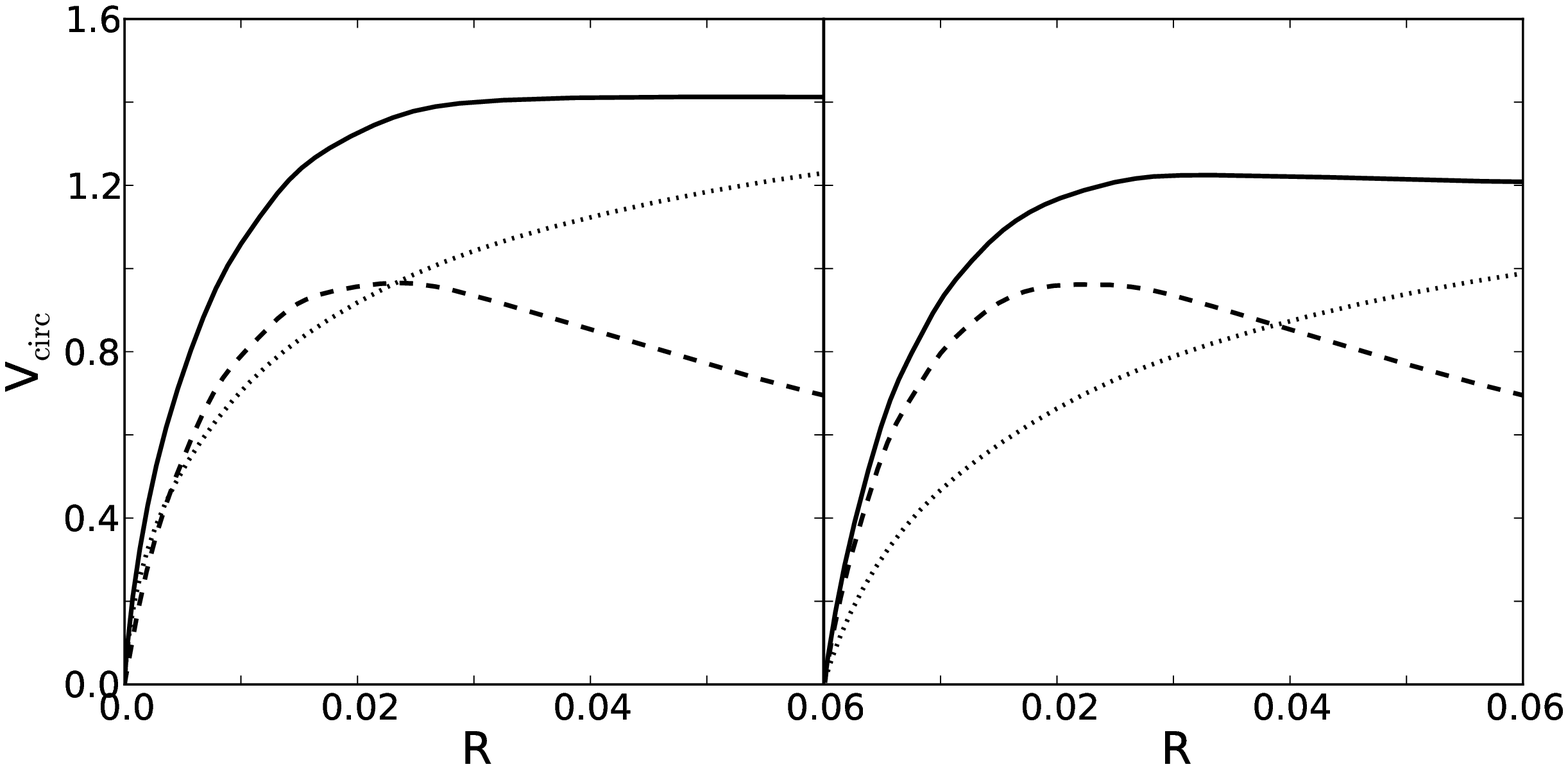}
\caption{ Initial circular velocity curves for the two
  simulations. Contributions from the disk and halo are shown as
  dashed and dotted lines, respectively. The solid line shows the
  total. Left: cuspy halo. Right: cored halo. The disk is the same
  in both simulations.}
\end{figure}


\section{Analysis}
To analyze orbits, we have adopted the use of a $k$means algorithm
\citep{lloyd82} orbit
classifier. Briefly, our $k$means
algorithm iteratively determines $k$ spatial centers of the set of $(x,y)$
apsis (local maxima in $r$) positions for a given orbit in the bar frame. In the bar frame,
apsides of trapped orbits will remain (within some tolerance) stationary
in spatial position. For example, in a 2:1 orbit the classifier will find the position of
$k=2$ centroids for the 
apsides, which will be located along the major axis of the bar. 
The orbital decomposition for the cuspy-halo system is shown in Figure 2. We find
three primary orbit families: (1) 2:1 orbits that make up the bar
quadrupole potential, (2) 
2:1$_\perp$ orbits that are perpendicular to the bar
, and (3) 4:1
orbits that align with the bar and comprise the bulk of the vertically
thickened structure. We classify the other orbits to be in the 'field'--
these orbits are overwhelmingly outside the influence of the bar and reside on
mostly circular orbits.
\begin{figure}[h]

\centering
\includegraphics[width=0.95\textwidth]{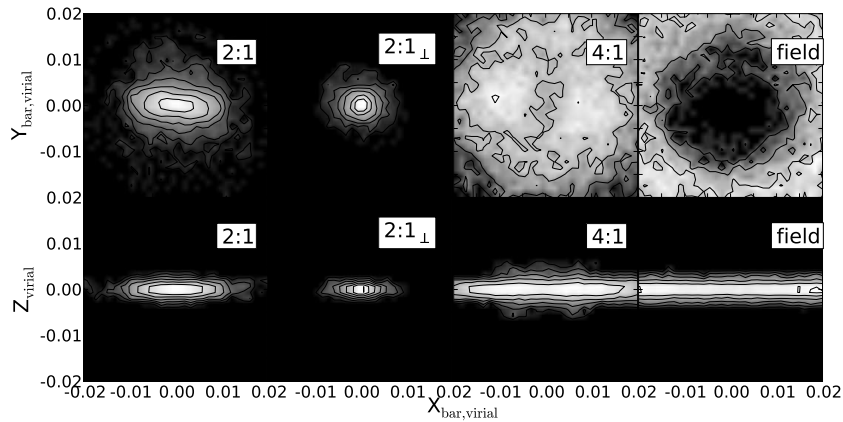}
\caption{Orbital decomposition for the cuspy halo run after bar formation. Time-integrated
orbits over the entire simulation are shown in grayscale. Contours are
spaced in 0.5 dex log density. Upper row: face-on view in the frame of
the bar. Lower row: edge-on view looking at the edge of the bar. Left:
2:1 orbits. Left center: 2:1$_\perp$ orbits. Right center: 4:1
orbits. Note the peanut-shaped thickening. Right: field stars, which
show no coherant structure and low density at $R<0.02$.}
\end{figure}

Both simulations result in the same three primary orbit
families. However, the fractional occupation of each family
differs. In the cusp simulation, the three family decomposition
returns fractional occupations of 
[2:1, 2:1$_\perp$, 4:1, field] = [0.226, 0.023, 0.311, 0.440], while the
core simulation returns [0.316, 0.056, 0.238, 0.390]. To first order, this gives a 
measure of the bar mass at late times by identifying orbits that
are trapped in the potential of the bar. In addition, separating the disk into distinct orbit families allows for the
analysis of angular momentum channels. In Figure 3 we plot the
fractional angular momentum (left, black axis) change for each orbit family during the
simulations. The 2:1 orbits that make up the bulk of the bar (as well
as the 2:1$_\perp$ orbits) quickly
shed the majority of their angular momentum in both simulations. Overlaid on the plot in gray is the total power in the
$m=2$ component of the potential expansion (right, gray axis), which is a measure of bar
strength. This gives some understanding of when the bar formation
epoch has ended ($T\approx0.70$ and $T\approx0.35$ for the cusp and core,
respectively). At this time, the angular momentum transfer becomes
roughly linear in 2:1 orbits. 4:1 orbits show a roughly linear decrease in angular
momentum over the course of the simulation. We
understand this as the catalyzation of the disk into trapped elevated
populations by the bar. 

An illuminating difference is
observed in the field populations of the simulations. In the cusp simulation,
the field stars have nearly constant angular momentum, as the trapped
orbits transfer angular momentum to the halo through resonant coupling
channels. In the cored simulation, the field stars gain the angular momentum
that the trapped orbits lost during the bar formation phase, as
the cored halo is unable to accept angular momentum at fractions of a
bar radius (as expected owing to the lower resonance density and
phase-space gradient in a core). In total, the halo accepts 10\% (5\%) of the disk angular momentum
in the cusp (core) simulation.
\begin{figure}[h]

\centering
\includegraphics[width=0.80\textwidth]{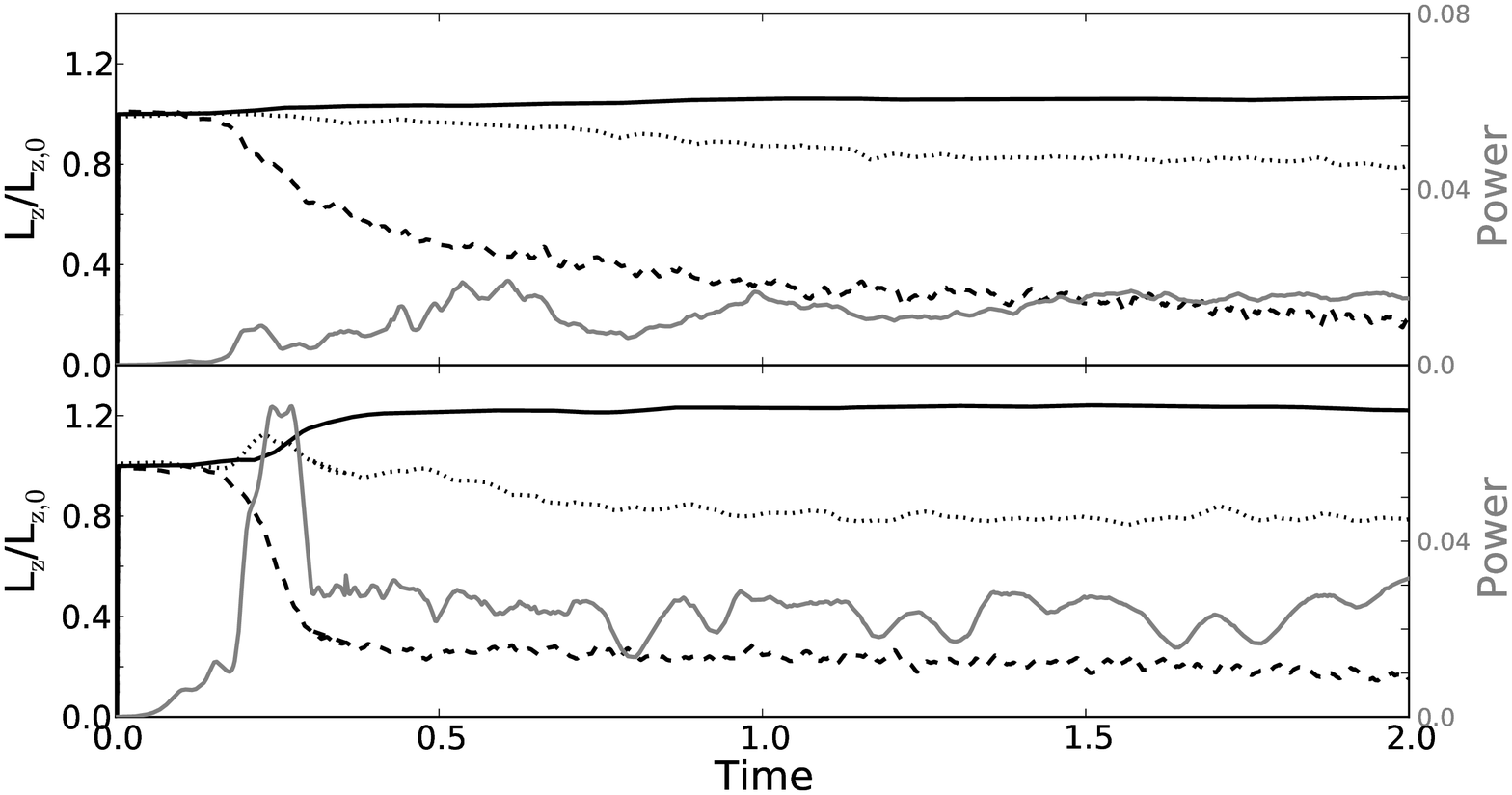}
\caption{Black (left axis): angular momentum relative to initial angular momentum
  for the (2:1 + 2:1$_\perp$), 4:1, and field orbits (dashed, dotted,
  and solid, respectively). Gray (right axis): $m=2$ power relative to $m=0$ for
  the potential realization. Upper panel: cuspy halo. Lower panel:
  cored halo. The cored halo buckles at $T=0.2\to0.3$, indicated by
  asymmetric terms in the potential expansion.}
\end{figure}
\section{Conclusion}

The main findings of this ongoing work are twofold: (1) we characterize the contrast between two modes of secular bar growth in
N-body simulations, one of which, the slow mode, deserves more
dedicated study and (2) present a method to accurately split the
simulations into dynamically distinct orbit populations that can be applied generally. We have begun to
isolate angular momentum transfer channels to illustrate the
variations in dynamical mechanisms between two bar formation modes, both of
which can result in a boxy or peanut-shaped bulge.

\acknowledgements This material is based upon work supported by the National Science
Foundation under Grant No. AST-0907951.

\bibliography{PetersenM}

\end{document}